# A note on the variability of V538 Cassiopeiae

Gustav Holmberg
Karl XI-gatan 8A
SE-222 20 Lund
Sweden
e-mail: 5063.gustav.holmberg@gmail.com

**Abstract**

CCD observations of V538 Cas have been made on nine nights during three weeks using the AAVSO Bright Star Monitor. No significant variations were found.

**1. Discovery**

Weber (1958a) discovered variations in the star BD 60°201 (HD 7681, HIP 6084, RA 01 18 07.2 Dec +61 43 04 (J2000)) by analysing plates taken between 1942 and 1958. The star was found to vary between 9.0 and 9.6 photographic magnitude and Weber, who did not publish a light curve, suspected it could be an eclipsing variable of the Algol type. A further study of the star was made by Häussler (1974) who used 135 patrol camera plates to find the star varying irregularly between 9.44 and 10.01 photographic magnitude. He classified it as an irregular variable star, type Isb.

On the basis of these studies, the star was designated V538 Cas. The entry in the *General Catalogue of Variable Stars* (GCVS; Kholopov et al., 1985) of the star has type Isb and gives the magnitude range as 9.4-10.6 photographic magnitude, and spectral type K5III. Later observations have re-classified the spectrum as M0 III (Henry et al 2000).

**2. New observations**

V538 Cas is, at 7.7V, a bright star. It is thus a suitable object for the AAVSO Bright Star Monitor (BSM; AAVSO 2009). This instrument aims at filling a niche in current photometric telescope ecology where not many CCD-equipped telescopes are operating: there are interesting stars that are difficult to observe using most modern telescopes because the stars are too bright and risk saturating the CCD detector. Finding suitable comparison stars for bright stars in telescopes with small fields of view can also be a challenge. Several of the current photometric surveys, such as NSVS and ASAS, only measure stars fainter than about 8th magnitude, giving much room for work for an instrument such as the Bright Star Monitor. The BSM is a 6 cm f/6.2 refractor with a SBIG ST-8XME CCD with a field of view of 84'x127', operated for AAVSO by Tom Krajci at the Astrokolkhoz facility in New Mexico.

30 s (V) and 60 s (B) exposures of V538 Cas were obtained with the BSM on nine nights during three weeks in November and December 2011. The images were analysed using VPHOT (AAVSO 2011) and the brightness of the star measured relative to an ensemble of five stars. The results (table 1) shows very small or no significant variations during this time interval.

These observations are in line with other measurements from the post-photographic era. A photometric study by Henry et al found slight short-time scale variations on the order of a hundredth of magnitude (Henry et al. 2000). Hipparcos (Perryman et al. 1997) consistently measured V538 Cas around 7.75 with variations on the order of hundredths of a magnitude. TASS, The Amateur Sky Survey (2012), observed the star on four occasions during three weeks, finding a constant brightness, also in the vicinity of 7.75 (see web links in the references section). Together with the new observations reported here, this leads the present author to conclude that V538 Cas, at least today, does not show the type of variations once attributed to it.

## 3. Discussion

Although not the primary purpose of this short communication, an attempt will also be made to try to resolve the discrepancy between data showing rapid variations on the order of 0.6 magnitude or more, and data showing no or very small variations in the star known as V538 Cas.

One possibility is of course that this star has changed its behaviour since the mid-20th century. Another is that Weber's initial guess, that this is an EA star, is correct and that further eclipses have not been covered in the observations made with the BSM, Hipparcos, Tass and by Henry et al. But can there be another explanation?

A similar discrepancy found in the literature may provide a clue. Weber, also in 1958, published his discovery of a cepheid in the open cluster NGC 7789, varying between photographic magnitude 11.2-12.2. The discovery was confirmed by another observer, Romano (Weber 1958b). Cepheids in open clusters have great astrophysical importance, and Weber's finding was therefore followed up. Burbidge and Sandage found no variations at all in the star in their photographic photometry of the cluster using the 100 inch Hooker telescope (Burbidge and Sandage 1958). Furthermore, Starrfield's photoelectric monitoring of the star with a 24 inch reflector at Lick during three hours per night during three nights found no variations in the star. Photographic photometry on a series of plates taken with the 20 inch Carnegie astrograph at Lick gave a similar result: constant brightness (Starrfield 1965). Measurements on plates from the Harvard College Observatory plate archives gave a similar result (Janes 1977).

Thus, we have two cases in which Weber found variations of quite large amplitude, both of which were first confirmed by another observer using similar type of cameras (Weber used small-scale photographic cameras) that was not confirmed by later observers. Starrfield, in trying to resolve this discrepancy, pointed out that a possible solution could be the small plate scale of Weber's camera; the suspected variable star was therefore imperfectly separated from another star, and seeing variations could produce a spurious impression of variability (Starrfield 1965). Perhaps this or some other photographic effect can account for the variations found in V538 Cas by Weber and Häussler.

## 4. Acknowledgment

The author wishes to thank AAVSO Director Arne A. Henden for his generous support. This research has made use of the SIMBAD database (CDS 2007), operated at CDS, Strasbourg, France.

Table 1: BSM measurements of V538 Cas in November-December 2011

| JD | Magnitude | filter |
|---|---|---|
| 2455881.85384 | 9.422 | B |
| 2455881.85444 | 7.696 | V |
| 2455882.80720 | 9.433 | B |
| 2455882.80782 | 7.715 | V |
| 2455888.78655 | 9.422 | B |
| 2455888.78714 | 7.713 | V |
| 2455889.68596 | 9.455 | B |
| 2455889.68656 | 7.699 | V |
| 2455892.65948 | 9.429 | B |
| 2455892.66008 | 7.715 | V |
| 2455893.65671 | 9.425 | B |
| 2455893.65730 | 7.718 | V |
| 2455894.67079 | 9.435 | B |
| 2455894.67139 | 7.702 | V |
| 2455895.65212 | 9.404 | B |
| 2455895.65272 | 7.697 | V |
| 2455896.64828 | 9.399 | B |
| 2455896.64888 | 7.706 | V |